\documentclass[aps,prx,floatfix,twocolumn,superscriptaddress]{revtex4-2}
\usepackage{color}

\usepackage{graphicx,
            amsmath,
            braket,
            xcolor,
            xspace,
            lipsum}
\usepackage{braket}

\begin{document}

\title{High-fidelity universal gates in the $^{171}$Yb ground state nuclear spin qubit}

\author{J.~A.~Muniz}
\thanks{J.~A.~M., M.~S.~and D.~T.~S.~contributed equally to this work.}
\author{M.~Stone}
\thanks{J.~A.~M., M.~S.~and D.~T.~S.~contributed equally to this work.}
\author{D.~T.~Stack}
\thanks{J.~A.~M., M.~S.~and D.~T.~S.~contributed equally to this work.}

\author{M.~Jaffe}
\author{J.~M.~Kindem}
\author{L.~Wadleigh}
\author{E.~Zalys-Geller}
\author{X.~Zhang}
\author{C.-A.~Chen}
\author{M.~A.~Norcia}
\author{J.~Epstein}
\author{E.~Halperin}
\author{F.~Hummel}
\author{T.~Wilkason}
\author{M.~Li}
\author{K.~Barnes}
\author{P.~Battaglino}
\author{T.~C.~Bohdanowicz}
\author{G.~Booth}
\author{A.~Brown}
\author{M.~O.~Brown}
\author{W.~B.~Cairncross}
\author{K.~Cassella}
\author{R.~Coxe}
\author{D.~Crow}
\author{M.~Feldkamp}
\author{C.~Griger}
\author{A.~Heinz}
\author{A.~M.~W.~Jones}
\author{H.~Kim}
\author{J.~King}
\author{K.~Kotru}
\author{J.~Lauigan}
\author{J.~Marjanovic}
\author{E.~Megidish}
\author{M.~Meredith}
\author{M.~McDonald}
\author{R.~Morshead}
\author{S.~Narayanaswami}
\author{C.~Nishiguchi}
\author{T.~Paule}
\author{K.~A.~Pawlak}
\author{K.~L.~Pudenz}
\author{D.~Rodr\'iguez~P\'erez}
\author{A.~Ryou}
\author{J.~Simon}
\thanks{Department of Physics and Department of Applied Physics, Stanford University, Stanford, California 94305, USA}
\author{A.~Smull}
\author{M.~Urbanek}
\author{R.~J.~M.~van de Veerdonk}
\author{Z.~Vendeiro}
\author{T.-Y.~Wu}
\author{X.~Xie}
\author{B.~J.~Bloom \\ Atom Computing, Inc.}\email{bbloom@atom-computing.com}

\begin{abstract}
\noindent Arrays of optically trapped neutral atoms are a promising architecture for the realization of quantum computers. In order to run increasingly complex algorithms, it is advantageous to demonstrate high-fidelity and flexible gates between long-lived and highly coherent qubit states. In this work, we demonstrate a universal high-fidelity gate set with individually controlled and parallel application of single-qubit gates and two-qubit gates operating on the ground-state nuclear spin qubit in arrays of tweezer-trapped $^{171}$Yb atoms.  We utilize the long lifetime, flexible control, and high physical fidelity of our system to characterize native gates using single and two-qubit Clifford and symmetric subspace randomized benchmarking circuits with more than 200 CZ gates applied to one or two pairs of atoms. We measure our two-qubit entangling gate fidelity to be 99.72(3)\% (99.40(3)\%) with (without) post-selection. In addition, we introduce a simple and optimized method for calibration of multi-parameter quantum gates. These results represent important milestones towards executing complex and general quantum computation with neutral atoms.  

\end{abstract}
\maketitle

\section{Introduction}\label{introduction}

Error-corrected quantum computation requires the ability to perform high-fidelity gate operations and readout on large numbers of physical qubits.  Toward this end, platforms utilizing individually controlled neutral atoms have recently demonstrated techniques to assemble arrays of over one thousand atomic qubits \cite{norcia2024iterative,pichard_rearrangement_2024,manetsch_tweezer_2024}, as well as mid-circuit measurement \cite{singh2023mid,Deist2022mid,graham2023mid,norcia2023midcircuit,lis2023mid,bluvstein2023logical}.  High-fidelity single-qubit gates with arbitrary local control have been demonstrated in atoms featuring hyperfine \cite{bluvstein2023logical} and nuclear spin qubits \cite{barnes2022assembly}.  Two-qubit gates with fidelity above 99\%, the most commonly cited threshold for the surface code~\cite{stephens_fault-tolerant_2014}, have been demonstrated in hyperfine~\cite{levine_parallel_2019,evered_high-fidelity_2023}, optical~\cite{madjarov2020high}, and metastable nuclear spin qubits~\cite{peper_spectroscopy_2024}.

Among the different optically trapped neutral atom platforms, ground-state nuclear spin qubits feature long coherence times due to a high degree of insensitivity to environmental perturbations such as trap light shifts and magnetic fields, as well as a near-infinite lifetime with respect to decay \cite{barnes2022assembly, jenkins2022yb}. Compared to metastable nuclear spin or optical qubits (which suffer from a relatively short lifetime due to trap Raman scattering) and hyperfine qubits in alkali atoms (which feature a relatively high sensitivity to trap light shifts), nuclear spin ground state qubits are well-suited for combining high-fidelity gates with atomic rearrangement. This paradigm facilitates flexible connectivity, thereby enabling the implementation of error correction schemes that efficiently utilize physical qubits~\cite{xu2024constant, hong2024long, bluvstein2023logical}.

Universal, gate-based quantum computation generally consists of a maximally entangling two-qubit gate (e.g. CNOT~\cite{cirac_quantum_1995}, XX~\cite{sorensen_quantum_1999} , CZ~\cite{urban_observation_2009}) and individually controlled single-qubit gates. In neutral atom systems, such single-qubit gates can be difficult in practice for closely spaced trapped qubits. This problem has recently motivated the use of global single-qubit operations for subspace benchmarking that can characterize many important gate errors in quantum systems~\cite{baldwin_subspace_2020, evered_high-fidelity_2023,tsai_benchmarking_2024}. However, for a more accurate measurement of overall gate fidelity, individually addressable single-qubit gates allow one to use two-qubit Clifford benchmarking~\cite{emerson_scalable_2005,dankert_exact_2009} that is not immune to specific errors.

\begin{figure*}[htb]
		\includegraphics[width=2.0\columnwidth]{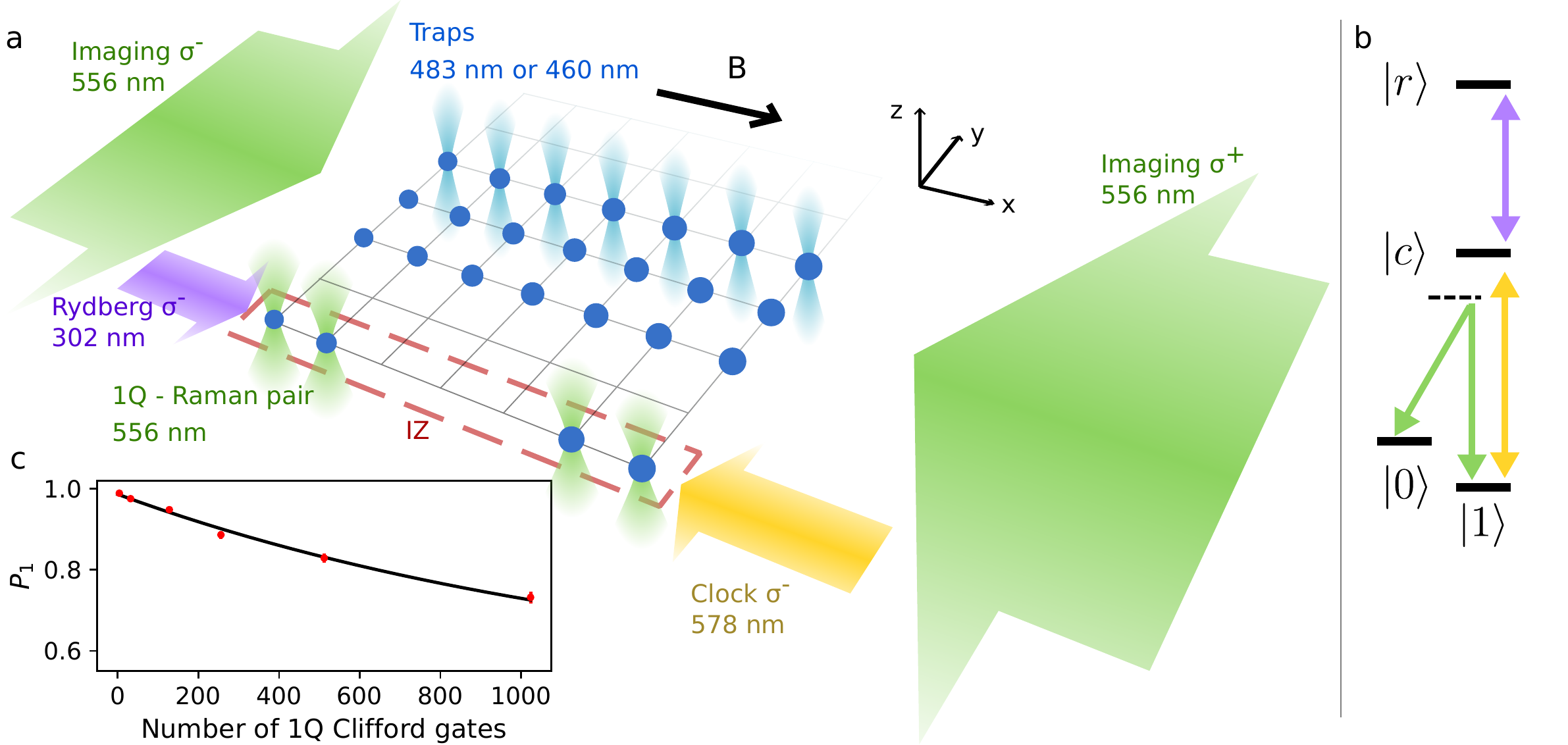}
		\caption{System description and single-qubit rotations. (a) Diagram showing an array of trapped atoms (blue circles) and the interaction zone (red box), where spatially selective global addressing for our two-qubit gates occurs. All IZ-global addressing beams, propagating along the $x$ direction, relevant for readout (green), clock (yellow) and Rydberg (purple), with their respective polarizations indicated, are shown. Individual Raman pair beams used for single-qubit gates propagate along $z$ and are linearly polarized along $x$ and $y$, respectively. Two-qubit gates are realized simultaneously in distant atomic pairs. Magnetic field $B$ (518~Gauss) points along the $x$ direction. (b) Level diagram showing the relevant transitions and atomic levels involved in our gates: $\ket{0}$ and $\ket{1}$ nuclear spin qubit states, clock state $\ket{c}$, and Rydberg state $\ket{r}$. Virtual level for the Raman pair is shown as a dashed line. Lasers transferring population between states are shown. (c) Clifford RB decay averaged over the seven IZ sites. Each depth is an average over 5 random 1Q Clifford circuits and 75 repetitions of each circuit. Error bars represent $1\sigma$ confidence interval.}
		\label{fig:fig1}
\end{figure*}

In this work, we demonstrate manipulation of quantum states in the ground-state nuclear spin of $^{171}$Yb atoms through individually controlled single-qubit gate operations and two-qubit operations based on sequential state-selective coherent excitation to a long-lived clock state and Rydberg state. We estimate a controlled-Z (CZ) fidelity of 99.72(3)\% (99.40(3)\%) with (without) post-selection from two-qubit Clifford Randomized Benchmarking (CRB) experiments, and a single-qubit CRB fidelity of 99.963(2)\%. Using a benchmark sequence that is insensitive to single-qubit phases (relevant to situations where such phases can be canceled between pairs of gates by using echo techniques as in \cite{evered_high-fidelity_2023,bluvstein2023logical}) and anti-symmetric errors, we infer a CZ gate fidelity of 99.84(6)\% (99.56(5)\%) with (without) post-selection. The combination of these spatially selective high-fidelity gates with the previously demonstrated continuous loading \cite{norcia2024iterative} and mid-circuit measurement \cite{norcia2023midcircuit} is expected to lead to new demonstrations of quantum error correction and complex circuits among many physical qubits, as we have recently shown in \cite{2024AtomQEC}.

\section{Qubit addressing}\label{system}

Our platform consists of an array of optically trapped single $^{171}$Yb atoms, previously described in Refs. \cite{norcia2023midcircuit,norcia2024iterative}, where the nuclear spin ground states $^1$S$_0$, $m_f = -1/2$ ($m_f = 1/2$) are used to encode the $\ket{0}$ $(\ket{1})$ states of our qubit, as shown in Figure \ref{fig:fig1}(b).  In this work, we operate with a modality specifically designed to isolate the performance of gate operations. Single and two-qubit gates are performed in the \textit{science} optical tweezers formed with 460~nm light (which provides state-insensitive trapping for the $^1$S$_0 \leftrightarrow ^3$P$_0$ optical clock transition used in our two-qubit gates \cite{norcia2023midcircuit,hohn2023state}), while state preparation and state-sensitive, nondestructive readout are performed in the \textit{reservoir} 483~nm tweezers (which provides state-insensitive trapping for the $^1$S$_0 \leftrightarrow ^3$P$_1$ inter-combination line). In order to enhance the data-rate while limiting the potential for inhomogeneity between sites, we perform two-qubit gates within one or two pairs of traps in a single row of the array, that is refilled from a larger reservoir of atoms after readout (see supplementary material SM for further details). We call this row the interaction zone (IZ), as sketched in Figure \ref{fig:fig1}(a).

\section{Single-Qubit Gates}

The single-qubit addressing scheme employed in this work (similar to that developed in Ref.~\cite{barnes2022assembly}) enables local and parallel control of the pulse area and phase of rotations applied to the nuclear spin qubits of multiple atoms in the IZ simultaneously. Here the Raman beams are red detuned by 5~GHz of the $^1$S$_0 \leftrightarrow ^3$P$_1,~F=1/2$ transition (see Figure \ref{fig:fig1}(a)-(b)), resulting in a two-photon Raman Rabi rate of $\sim 2\pi \times 7$~kHz. The phase of the applied gate is set by the differential phase of the two Raman beams, and can be controlled arbitrarily.

The single-qubit (1Q) gate set used here consists of $Z_{\pi/2}$ and $X_{\pi/2}$ operations, where $Z_{\pi/2}$ operations are performed virtually via frame-tracking and used to update the phase of the next applied $X_{\pi/2}$ pulse \cite{McKay2017Zgates}. We characterize the performance of these gates using Clifford Randomized Benchmarking (CRB) experiments~\cite{knill_randomized_2008,Magesan2011RB,Magesan2012gates}, yielding an average fidelity of 99.963(2)\% per Clifford gate, averaged over the seven IZ sites. Figure~\ref{fig:fig1}(c) shows a typical averaged RB curve for seven IZ sites, with the CRB circuit executed in parallel among the IZ atoms. 

Calibration of the $X_{\pi/2}$ gate requires setting the pulse area by tuning the product of Rabi rates associated with the individual Raman beams, and zeroing differential light shifts by tuning the intensity ratio of the two Raman beams. For both of these calibrations, we rely on pulse sequences heavily inspired by Robust Phase Estimation (RPE) methods~\cite{Kimmel2015RPE}. Additionally, we perform local beam alignment calibrations via single beam AC-Stark shift measurements to align each set of Raman beams to the atoms~\cite{bluvstein2023logical}. Typical gate performance is limited by quasi-static drifts in alignment and beam intensities, not fundamental processes such as intermediate state scattering. All calibrations and corrections are applied to individual qubits (see SM).

\section{Two-qubit gates}\label{2q}

Two-qubit (2Q) gates are performed by state-selectively exciting pairs of atoms to high-lying Rydberg states via a two-step process to apply a symmetric CZ gate \cite{levine_parallel_2019}.  
In contrast to previously demonstrated two-photon excitation schemes, where the drive lasers are applied simultaneously and detuned from a short-lived intermediate state~\cite{wilk2010entanglement,graham_rydberg-mediated_2019, levine_parallel_2019,ma_universal_2022}, we operate with sequential, resonant excitation to and from a long-lived intermediate state \cite{madjarov2020high}. This has several key advantages. It allows us to maximize the Rabi frequency to the short-lived Rydberg state given power constraints, which in turn reduces the effects of Rydberg decay. Further, by using a long-lived intermediate state, scattering from this state is reduced as well. Finally, because the excitation to the intermediate state is relatively slow, moderate differential light shifts on the narrow transition can be used to prevent atoms from participating in a gate, providing opportunities for site-selective addressing while using global gate lasers~\cite{norcia2023midcircuit}.

Excitation to and from the metastable clock state $\ket{c}=\ket{^3\text{P}_0,~m_f=-1/2}$ that forms the intermediate state of our sequential excitation scheme is performed via a $X^{\mathrm{clk}}_{\pi}$ shelving pulse, designed and calibrated to transfer as much population as possible between $\ket{1}$ and $\ket{c}$. A combination of frequency- and polarization-selectivity provides state-selectivity for this excitation process, and thus for the two-qubit gate. From the clock state, we apply a pulse of ultraviolet (UV), 302~nm, $\sigma^-$-polarized light to drive the $\ket{c} \leftrightarrow \ket{r} = \ket{65~^3\text{S}_1,~F=3/2,~m_f=-3/2}$ transition (see Figure \ref{fig:fig1}(b)). The UV pulse phase and amplitude profile are chosen to ensure that every atomic pair returns to its initial state after the pulse, while pairs of neighboring clock atoms acquire an additional $\pi$ phase shift due to the Rydberg blockade mechanism \cite{Jaksch2000RydGate}. Finally, atoms are returned to $\ket{1}$ with a second $X^{\mathrm{clk}}_{\pi}$ pulse, having acquired a conditional phase that implements a CZ gate.  

During the application of our two-qubit gates, the atoms are trapped in 460~nm optical tweezers within the IZ with a trap frequency $\omega_\mathrm{x}/(2\pi) = 50$~kHz. Non-participating reservoir atoms are maintained in $\ket{0}$, and do not couple to $\ket{c}$. 

\subsection{Clock Shelving}\label{clock}
To ensure optimal performance of the two-qubit gate, it is crucial to minimize population and phase errors arising from the clock shelving and unshelving pulses. The clock pulses are applied to a Doppler-sensitive (single-photon) transition, and are relatively slow ($\Omega_{\mathrm{clk}}/(2\pi) \approx 7$~kHz) compared to the trap frequency, making them sensitive to a specific set of errors: finite atomic temperature leads to a spread in Rabi frequencies between motional states \cite{Wineland1979cooling}, and atoms can be coupled to other motional states of the trap. Laser phase and amplitude noise near the Rabi frequency \cite{finkelstein2024universal,ball2016oscillator,Jiang2023}, as well as trap-induced decay from the clock state \cite{Dorscher2018Raman} can also degrade performance. Finally, quasi-static errors in clock laser detuning can lead to qubit phase shifts during the gate.  

\begin{figure}[htb]
		\includegraphics[width=\columnwidth]{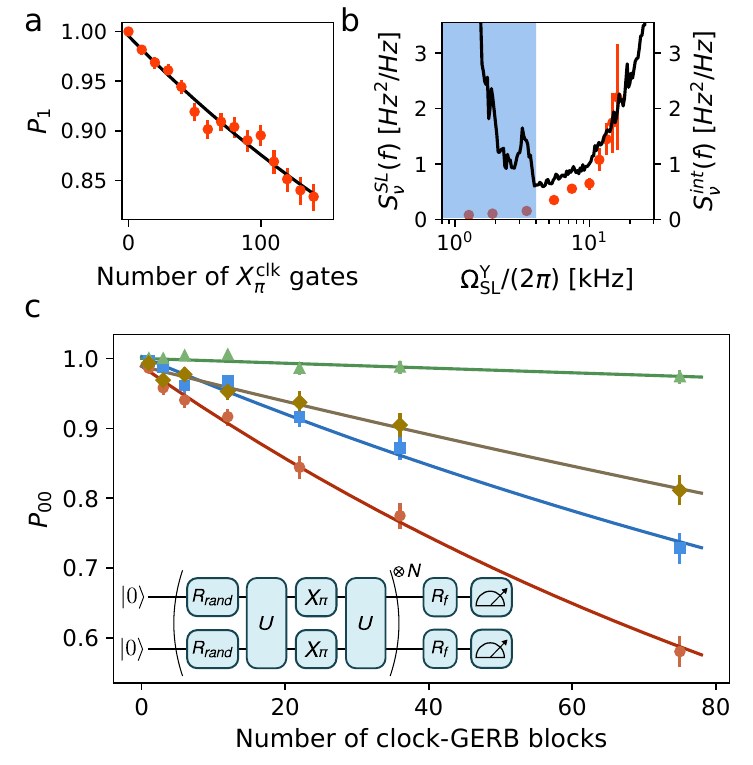}
		\caption{Clock benchmark. (a) Probability $P_1$ of measuring an atom in state $\ket{1}$, after repeated application of SCP $X^{\mathrm{clk}}_{\pi}$ pulses on the IZ, for initial state $\ket{1}$. Here we measure 99.860(7)\% shelving fraction per pulse, per atom. (b) Spin locking measurement for the $X^{\mathrm{clk}}_{\pi/2}-Y^{\mathrm{clk}}(T)-X^{\mathrm{clk}}_{-\pi/2}$ circuit, where $Y^{\mathrm{clk}}(T)$ is a pulse along the geometrical Y axis in the $\ket{1}$ - $\ket{c}$ Bloch sphere with duration $T$ and Rabi rate $\Omega^{\mathrm{Y}}_{\mathrm{SL}}$. For each $\Omega^{\mathrm{Y}}_{\mathrm{SL}}$, we infer the laser frequency noise $S^{\mathrm{SL}}_{\nu} (f)$ from the temporal decay in coherence. We also show the interferometer laser frequency noise $S^{\mathrm{int}}_{\nu} (f)$ (black). Data below 4~kHz is heavily impacted by acoustic noise present in the interferometer (blue shaded region). (c) Pre-selected (red circles), post-selected on atoms remaining in the qubit subspace (blue squares), pair leakage (brown diamonds) and pair survival (green triangles) for a clock-GERB experiment ($U = X^{\mathrm{clk}}_{\pi}X^{\mathrm{clk}}_{\pi}$) measured in the two atom basis. Solid lines are fits to $ap^x+b$, where $x$ is the number of GERB blocks, and $b=1/9$ for the pre-selected curve, $b=1/4$ for the post-selected in qubit subspace curve, and $b=0$ for the loss and leakage curves (see SM Appendix ~\ref{analysis}). Error bars represent $1\sigma$ confidence intervals. Inset representing U-GERB sequence as described in the main text.}
		\label{fig:fig2}
\end{figure}

In order to mitigate the effects of atomic motion, we cool the atoms near their motional ground state along the direction of the clock laser. For cooling, we use 3D gray molasses to cool atoms to $\bar{n} = 0.25(10)$ along the $x$-direction \cite{lis2023mid}. With a Lamb-Dicke parameter $\eta = 0.26$, this would limit the clock-shelving fidelity to $< 99.8\%$~\cite{wineland_experimental_1998} for a single $\pi$ pulse. With two clock shelving pulses per qubit per CZ gate, clock shelving errors would limit CZ fidelity significantly. To further reduce sensitivity to any effect that causes spread in Rabi rates, and atomic temperature, we employ shaped composite pulses (SCPs) \cite{Levitt1986CPs}. The smooth pulse shape reduces unwanted frequency components that can induce motional state changing transitions, and the composite pulse is designed to be robust to pulse-area errors. For a detailed error budget, see the SM (Appendix \ref{gate_simulator}).

There are significant drawbacks to using SCPs. Pulses are typically longer than a square pulse, making them more susceptible to laser frequency noise, Raman scattering, and quasi-static detunings. To minimize these effects, we have found that a Blackman shaped $Y^{\mathrm{clk}}_{\pi/2}-X^{\mathrm{clk}}_{\pi}-Y^{\mathrm{clk}}_{\pi/2}$ pulse offers enough robustness, while balancing the impact from frequency noise \cite{LevittXYXPulse}. We refer to this pulse as the $YXY^{\mathrm{clk}}$ pulse and from now on we will assume that the $X^{\mathrm{clk}}_{\pi}$ operation is a $YXY^{\mathrm{clk}}$ pulse. Additional effects that affect clock shelving fidelity, but have smaller impact, can be found in the SM.

In Figure~\ref{fig:fig2}(a), we measure the shelving fidelity of our $X^{\mathrm{clk}}_{\pi}$ pulses for atoms starting on state $\ket{1}$. Typical shelving fidelities per pulse exceed 99.85\% in these conditions. Most of the error is population left in $\ket{c}$, and a smaller loss or decay to the other ground state due to Raman scattering of the 460~nm light. Each Blackman-shaped $\pi$-clock pulse lasts 130~$\mu$s, and is calibrated using RPE techniques. We characterize the laser frequency noise on an optical self-heterodyne fiber interferometer setup \cite{Kefelian2009FiberInterferometer}, and with spin locking atomic measurements that map laser frequency noise into changes of atomic coherence \cite{finkelstein2024universal,tsai_benchmarking_2024}. The frequency range of interest spans from a few kHz to 20 kHz, where typical laser locks have limited gain. We find that the interferometer reports a larger frequency noise than the one inferred from the atomic measurement, as shown in Figure \ref{fig:fig2}(b). We attribute this disagreement to a combination of excess acoustic noise and limitations on the noise floor of the interferometer, therefore we bound the contribution of laser frequency noise to the one extracted in the spin locking experiments. Simulations presented in the SM incorporate laser frequency noise to estimate the shelving fraction.

To characterize the performance of clock pulses (and later, two-qubit gates) on arbitrary nuclear spin qubit states, we adopt a strategy similar to the one described in \cite{evered_high-fidelity_2023}. This protocol consists of initializing atom pairs in $\ket{00}$, and applying $N$ blocks each containing (i) a common random Haar-distributed 1Q rotation $R_{\mathrm{rand}}$ on both qubits, (ii) a two-qubit unitary $U$ on the atomic pair, (iii) an echo pulse ($X_{\pi}$) on the qubit space, and (iv) an additional application of $U$, as sketched in Figure \ref{fig:fig2}(c) inset. Each block uses a different $R_{\mathrm{rand}}$, and satisfies that $U-X_{\pi}-U$ does not create entanglement. After the $N$ blocks are applied, a deterministic 1Q rotation $R_{\mathrm{f}}$, pre-calculated under the assumption that $U$ is ideal, returns atoms to the $\ket{00}$ state. Readout is performed in the two-qubit computational basis. By progressively constructing our CZ gate from different $U$ gates, for instance identity, $X^{\mathrm{clk}}_{\pi}X^{\mathrm{clk}}_{\pi}$, and finally a full CZ gate, we can identify different error sources. We call this protocol $U$-global echo randomized benchmarking sequence ($U$-GERB). We note that unlike 2Q Clifford RB, this U-GERB protocol applies the same gates to atoms within a pair and so constitutes a symmetric subspace benchmark.

We measure the characteristic clock-GERB curve for $U = X^{\mathrm{clk}}_{\pi}X^{\mathrm{clk}}_{\pi}$ in Figure~\ref{fig:fig2}(c). We notice that this measurement is subjected to errors from the eight 1Q gates used. We characterize those in the case where $U$ is the identity operation and measure a contribution of 0.32(2)\% per atomic pair per GERB block. For this set, characteristic of a well-tuned system, we measure a pre-selected fidelity for the shelve-unshelve clock sequence (red circles) of 99.80(1)\% per pair, after removing the 1Q error. The post-selected fidelity of the clock shelve-unshelve sequence on atoms that remain on the qubit subspace at the end of the circuit is 99.94(2)\% (blue squares), which is mostly affected by decoherence between the optical and ground state qubits. The difference between the pre- and post-selected fidelities points to leakage (atoms that remained in $\ket{c}$), and loss. 

Leakage error represents population left in the clock state after each shelve-unshelve sequence ($U$). We are able to measure this contribution using an additional readout step preceded by an clock repumping state, that effectively brings all of the clock remaining clock population back to the ground state. By estimating the fraction of pairs that are present in this readout image, compared to the image that only reveal atoms present in the qubit subspace, we determine that clock leakage (brown diamonds) contributes a 0.13(2)\% infidelity in our clock-GERB sequence. Finally, using all of the readout steps we can determine the pair survival probability (green triangles) of 99.979(4)\% after each shelve-unshelve sequence (see SM Appendix ~\ref{clock_app}).

\subsection{Rydberg Gate}\label{rydberg}

\begin{figure}[htb]
    \centering
    \includegraphics[width=0.9\linewidth]{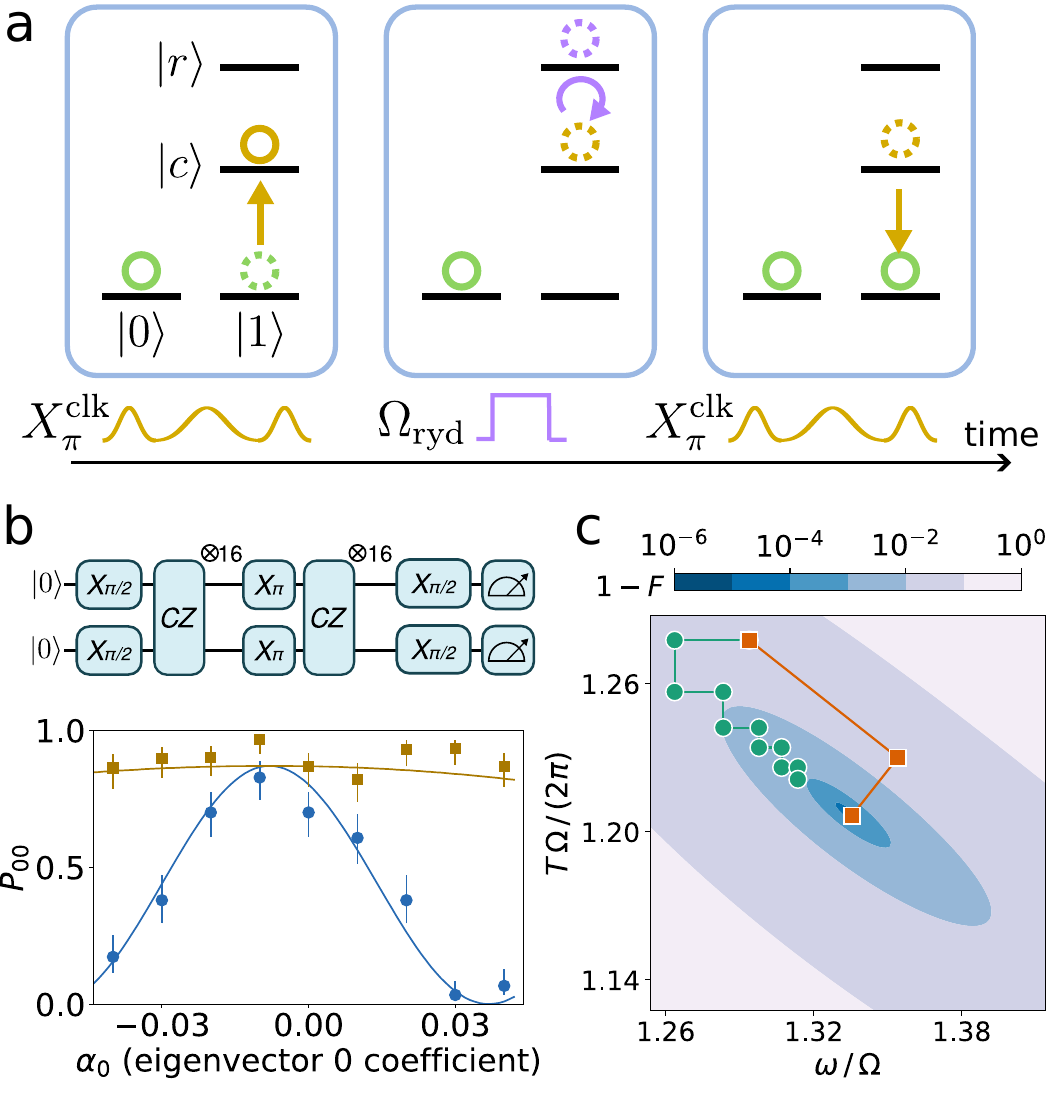}
    \caption{Two-qubit gate. (a) Schematic depiction of three-step gate. First, an $X^{\mathrm{clk}}_\pi$ pulse shelves qubit $\ket{1}$ into $\ket{c}$. Second, the clock state is coupled to a Rydberg state, with the laser amplitude and phase profile chosen to implement a CZ gate. Third, an unshelving $X^{\mathrm{clk}}_\pi$ pulse transfers population from $\ket{c}$ back to qubit $\ket{1}$. Each step is performed simultaneously on two participating atoms. (b) The CZ gate parameters are calibrated by a series of 1D scans along eigenvectors of the fidelity Hessian. After executing a circuit with repeated CZ gates (top), we measure population in $\ket{00}$ (blue circles) and pair survival (yellow squares). The eigenvector scan is $d\left(A, \omega/\Omega_{\mathrm{ryd}}, \phi, T\Omega_{\mathrm{ryd}}/(2\pi),\Delta/\Omega_{\mathrm{ryd}}\right)=\alpha_0\left(0.45, -0.19, 0.14, -0.056, 0.87\right)$. This eigenvector changes $\phi_{ent}$ while having little effect on survival. Solid curves are simulation results with fitted $x$ offset and $y$ scale. (c) Simulated optimization trajectory with repeated 1D scans of raw gate parameters (green circles) or eigenvector coefficients (orange squares). Eigenvector scans are decoupled and reach the minimum in a fixed number of steps, while raw parameter scans are not fully converged after many optimization rounds. For visual clarity, the optimization is performed over just two of the five phase parameters, $\omega$ and $T$. Contours represent the gate infidelity accounting only for calibration errors.}
    \label{fig:eig}
\end{figure}
Once clock shelving is complete, entangling gates are performed via the Rydberg blockade mechanism~\cite{saffman_quantum_2016}, by coupling atoms from the clock state to the $\ket{r}$ state with a single global beam at 301.9~nm (Figure~\ref{fig:eig}(a)). Our ultraviolet laser system enables Rydberg Rabi rates $\Omega_{\mathrm{ryd}} > 2\pi \times 15$~MHz, though usually a smaller Rabi rate is chosen to remain far below the Rydberg interaction energy $U/\hbar=2\pi\times 160$~MHz. For most Rydberg operations the 460~nm tweezers are kept on, as they provide a trapping potential for the $\ket{r}$ state due to the ion core polarizability~\cite{wilson2022trapping}, and have minimal impact on gate performance. The Rydberg state lifetime is 65(3)~$\mu$s at a typical trap depth (see SM).

CZ gates are implemented with an approximation of the time-optimal gate~\cite{jandura2022time}, using the sinusoidal phase parametrization of Ref.~\cite{evered_high-fidelity_2023}, $\phi(t) = A\cos(\omega t-\phi) + \Delta t$, with a square pulse of length $T$. Optimization over the phase parameters guarantees that the $\ket{01}$ ($\ket{10}$) states undergo nearly closed rotations through the $\ket{0r}$ ($\ket{r0}$) states, ideally leaving no population in the Rydberg state and picking up a single-qubit phase $\phi_{01}$ ($\phi_{10}$), while the $\ket{11}$ ideally leaves no population in $\ket{W}=(\ket{cr}+\ket{rc})/\sqrt{2}$ and picks up a different phase $\phi_{11}$. The latter may be decomposed into the sum of single-qubit and entangling phases as $\phi_{11}=\phi_{ent}+\phi_{01}+\phi_{10}$. Choosing gate parameters so that $\phi_{ent}=\pi$, the final unitary is $U_0=\text{diag}(1, e^{i\phi_{01}}, e^{i\phi_{10}}, -e^{i(\phi_{01}+\phi_{10})})$,
which can be converted to the ideal CZ gate $U_0=\text{diag}(1,1,1,-1)$ with virtual single-qubit $Z$ rotations.

While optimal phase profiles are readily obtained in simulation, experimental imperfections shift the optimum, requiring calibration of the control parameters. We implement an optimized calibration by diagonalizing the simulated Hessian matrix of the gate error. The resulting eigenvectors are nearly decoupled, so that near-optimal fidelity can be reached with a single 1D scan along each eigenvector (Figure~\ref{fig:eig}(b-c)). This strategy maintains the robustness and clarity of 1D scans, while greatly improving convergence compared to scans of the raw gate parameters (Figure~\ref{fig:eig}(c)).

The gate is calibrated using an echoed metric such as the population of the desired state after applying CZ$^N$-$X_{\pi}$-CZ$^N$, with $N$ gates on either side of the $X_{\pi}$ pulse, as shown in Figure \ref{fig:eig}(b) inset. The echo removes any dependence on the single-qubit phases, which are obtained separately by robust phase estimation. It is straightforward to incorporate effects like finite blockade and pulse rise time into the model, improving the initial estimate of optimal parameters and the decoupling of the eigenvectors.

\subsection{Two-Qubit Gate Benchmarking}\label{benchmarking}

Clifford gates are the basis for many error-correcting protocols envisioned for universal fault-tolerant quantum computation~\cite{gottesman_introduction_2009}. The Clifford group is sufficiently complex that most errors (even coherent errors) will be perfectly depolarized under the assumption of fixed Clifford- error channels associated with each Clifford gate. This allows one to make unbiased comparisons across different experimental platforms, even if the implementation of the Clifford operators may vary~\cite{gaebler_randomized_2012}. The ability to perform local and independent single-qubit operations is both important for use in quantum algorithms, and allows us to perform fidelity benchmarks that average over the full Hilbert space of the qubits, as opposed to only the symmetric subspace as has been the case in recent gate benchmarking approaches with tweezer-trapped neutral atoms~\cite{evered_high-fidelity_2023, ma2023high, peper_spectroscopy_2024, tsai_benchmarking_2024}. Physical error sources, such as decay from the Rydberg state, entangling phase, or single-qubit phase errors, will contribute to circuit errors with slightly different weights (see Appendix \ref{gate_simulator})~\cite{tsai_benchmarking_2024}.

\begin{figure}[htb]
    \centering
    \includegraphics[width=1.0\linewidth]{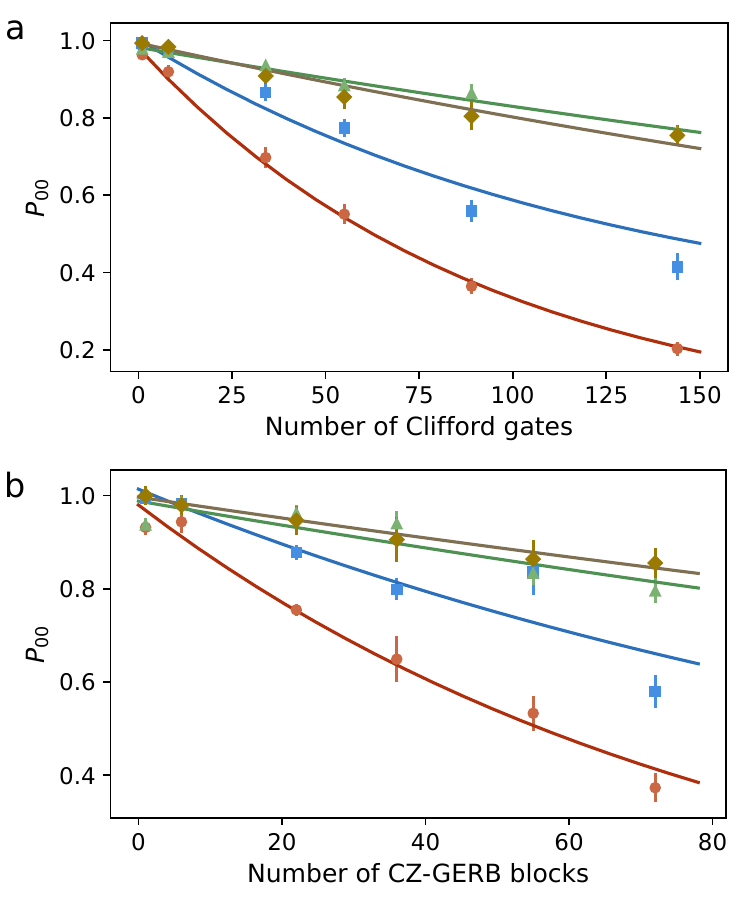}
    \caption{Two-qubit Clifford RB (a) and CZ-GERB (b) benchmarking, showing the probability of measuring atoms in their initial state $P_{00}$ as a function of gate depth. Blue squares (red circles) represent probability of returning to the initial two qubit state with (without) post-selecting on atoms that are in either nuclear spin qubit at the end of the circuit. Pair leakage (survival) measurements are shown with brown diamonds (green triangles). Solid curves are fits to the function $ap^x+b$, where $x$ is the circuit depth, and $b=1/4$ for the post-selected in qubit subspace curve, and $b=0$ for the pre-selected, leakage and survival curves (see SM). Error bars represent $1\sigma$ confidence intervals.}
    \label{fig:2q_benchmark}
\end{figure}

We implement two-qubit Clifford gates~\cite{Magesan2011RB} by performing the 1Q and 2Q gates described previously. For each realization of the experiment, twenty random quantum circuits with a given depth are generated with the Qiskit Experiments package~\cite{Quiskit2024} and executed 20 times per random circuit. Figure~\ref{fig:2q_benchmark}(a) summarizes our measurements. By fitting the data to an exponential decay curve, we measure an average two-qubit Clifford gate fidelity with (without) post-selection of 99.40(4)\% (98.93(4)\%). Accounting for the average number of native 2Q and 1Q gates per Clifford gate ($\approx 1.51$ and $\approx 4.36$ respectively), and assuming a depolarizing error model, we extract a CZ fidelity of 99.72(3)\% (99.40(3)\%) with (without) post-selection. Atom pair loss, the probability of losing at least one atom of the atomic pair, is measured to be 0.12(1)\% per CZ gate and is dominated by Rydberg state decoherence. 

We compare this 2Q CRB measurement to a CZ-GERB measurement, which represents a symmetric subspace benchmark of our CZ gate. In this case, the presence of echoes within each GERB block (similar to the method in reference~\cite{evered_high-fidelity_2023}) makes this circuit insensitive to single-qubit phase offsets, and more robust against quasi-static drifts. Using the CZ-GERB circuit we measure a fidelity of 99.84(6)\% (99.56(5)\%) with (without) post-selection on atoms remaining in the qubit subspace. We attribute a large fraction of the difference between the fidelity metrics to quasi-static errors of the clock laser detuning, which maps into single-qubit phase errors.

The excess pair loss (0.11(3)\%) measured in CZ-GERB (Figure \ref{fig:2q_benchmark}(b)) relative to clock-GERB (Figure \ref{fig:fig2}(c)) contributes half of the additional pre-selected infidelity measured between the two metrics, 0.20(2)\% and 0.44(5)\% respectively. Furthermore, clock state leakage is determined to be the same as the one measured in our clock-GERB data set, at 0.12(3)\% per gate, per pair for CZ-GERB data and 0.14(2)\% for the 2Q CRB set. Additionally, errors in entangling or single-qubit phases between consecutive CZ gates will lead to excess infidelity. Table \ref{tab:infidelities} summarizes the observed infidelities. We provide a more detailed final error budget supported by in-situ measurements and simulations in the SM, as well as details regarding the depth scans and fidelity extraction.

\begin{table}[htb]
    \centering
    \begin{tabular}{|p{2cm}|p{1.5cm}|p{1.5cm}|p{1.5cm}|p{1.5cm}|} 
        \hline
         & 1Q-GERB & Clock-GERB & CZ-GERB & 2Q CRB \\ [0.5ex] 
        \hline
        Pre-selected infidelity & 0.16(1)\% & 0.20(2)\% &  0.44(5)\% & 0.60(3)\% \\ 
        \hline
        Post-selected infidelity & 0.16(1)\% & 0.06(2)\% & 0.16(6)\% & 0.28(3)\%\\
        \hline
        Pair Leakage  & 0.00(1)\% & 0.13(2)\% & 0.12(3)\% & 0.14(2)\% \\
        \hline
        Pair Loss  & 0.00(1)\% & 0.021(4)\% & 0.13(3)\% & 0.12(1)\% \\
        \hline  
    \end{tabular}
    \label{tab:infidelities}
    \caption{Infidelity contribution, pair leakage and pair loss per CZ gate as measured by GERB and CRB. Clock-GERB, CZ-GERB, and 2Q CRB infidelities are corrected by the error of a single 1Q gate and the average number of 1Q gates in each circuit.}
    \label{tab:infidelities}
\end{table}

\section{Conclusions and Outlook}\label{disc}

In this work we have shown high-fidelity single- and two-qubit gates on ground state nuclear spin qubits.  While the demonstration of two-qubit gates was restricted to one or two pairs of interaction sites, the methods demonstrated here can also be applied to larger arrays with arbitrary connectivity by utilizing larger numbers of tweezer traps~\cite{norcia2024iterative} and coherent movement of atoms~\cite{bluvstein2022quantum}.  These extensions benefit from the long coherence time and insensitivity to light shifts of the ground-state nuclear spin qubit, though care must be taken to minimize atomic heating during movement.  Nuclear spin qubits also enable a complementary approach, where the narrow linewidth of the clock shelving transition allows the use of local light shifts to modify connectivity within static arrays of atoms (though connectivity in this approach is limited to nearby atoms).  By combining these techniques with our demonstrated mid-circuit measurement~\cite{norcia2023midcircuit} and continuous loading techniques~\cite{norcia2024iterative}, as well as erasure conversion provided by state-selective, non-destructive measurement~\cite{sahay_high-threshold_2023}, the high-fidelity and flexible gates demonstrated here are expected to enable the execution of complex error-corrected quantum circuits.  

As part of this work we extensively characterized our single- and two-qubit gates, identifying key areas for improvement, like clock laser quasi-static and fast frequency noise \cite{Li2022feedforward,finkelstein2024universal}, and further investigation of the complex Rydberg state manifold \cite{peper_spectroscopy_2024,hummel2024rydberg}. The high-fidelity measured in this report, coupled with advancements in efficient logical qubit encodings \cite{xu2024constant} and single-shot fault tolerant schemes \cite{Bombin2015Singleshot} enabled by all-to-all connectivity, place neutral atom arrays in an exciting position in the pursuit of practical quantum computing.

\appendix
\section*{Appendices}

\renewcommand{\thesubsection}{\Alph{subsection}}

Appendix~\ref{loading_imaging} describes our experimental setup and details regarding readout. Appendices ~\ref{1q}-\ref{2q_cals} describe details about our 1Q and 2Q gates. Details on data analysis are discussed in Appendix~\ref{analysis}. Our best understanding of our error budget is described in \ref{gate_simulator}.

\section{System Details} \label{loading_imaging}
Key aspects of our experimental system have been previously described in Refs.~\cite{norcia2023midcircuit,norcia2024iterative}. The lasers required to create both reservoir and science arrays, to rearrange atoms between reservoir trap sites, and the Raman beams needed for the single-qubit rotations are combined on a dichroic mirror stack. These beams are then delivered to the vacuum chamber via a high numerical aperture microscope objective (NA = 0.65, field of view = 0.5~mm). Another similar objective is placed at the other side of the vacuum chamber, and is used exclusively to collect 556~nm scattered light in order to perform low-loss, state-selective imaging of individual atoms. In this work, all state preparation and measurement operations are performed in optical tweezers, not a cavity-enhanced optical lattice as in \cite{norcia2024iterative}.  

The reservoir and science traps are each generated via SLM phase patterns \cite{nogrette_single-atom_2014} that are imaged on a microscope objective. The arrays used for this work consist of 84 trapping sites, distributed in 12 rows and 7 columns. The top row of each array, which we call the interaction zone (IZ), is displaced from the rest of the array by 6~$\mu$m. The distance between sites in the IZ row is 3~$\mu$m. The reservoir traps have radial trap frequencies (along the $x$ and $y$ directions) of $\omega_{\mathrm{res}}/(2\pi) = 110$~kHz, while the science trap frequency is $\omega_{\mathrm{sci}}/(2\pi)=50$~kHz. Both arrays are spatially matched using camera measurements on a lower numerical-aperture monitoring system, and by realizing atomic transfer experiments to calibrate any systematic offsets. To transfer atoms from one array to another, we linearly ramp up the power of one potential as we ramp down the other over 2~ms. All gates used in this work are performed in the science array IZ, while imaging, cooling, and state preparation are performed in sites of the reservoir array at the same locations.

\begin{figure}[htb]
		\includegraphics[width=\columnwidth]{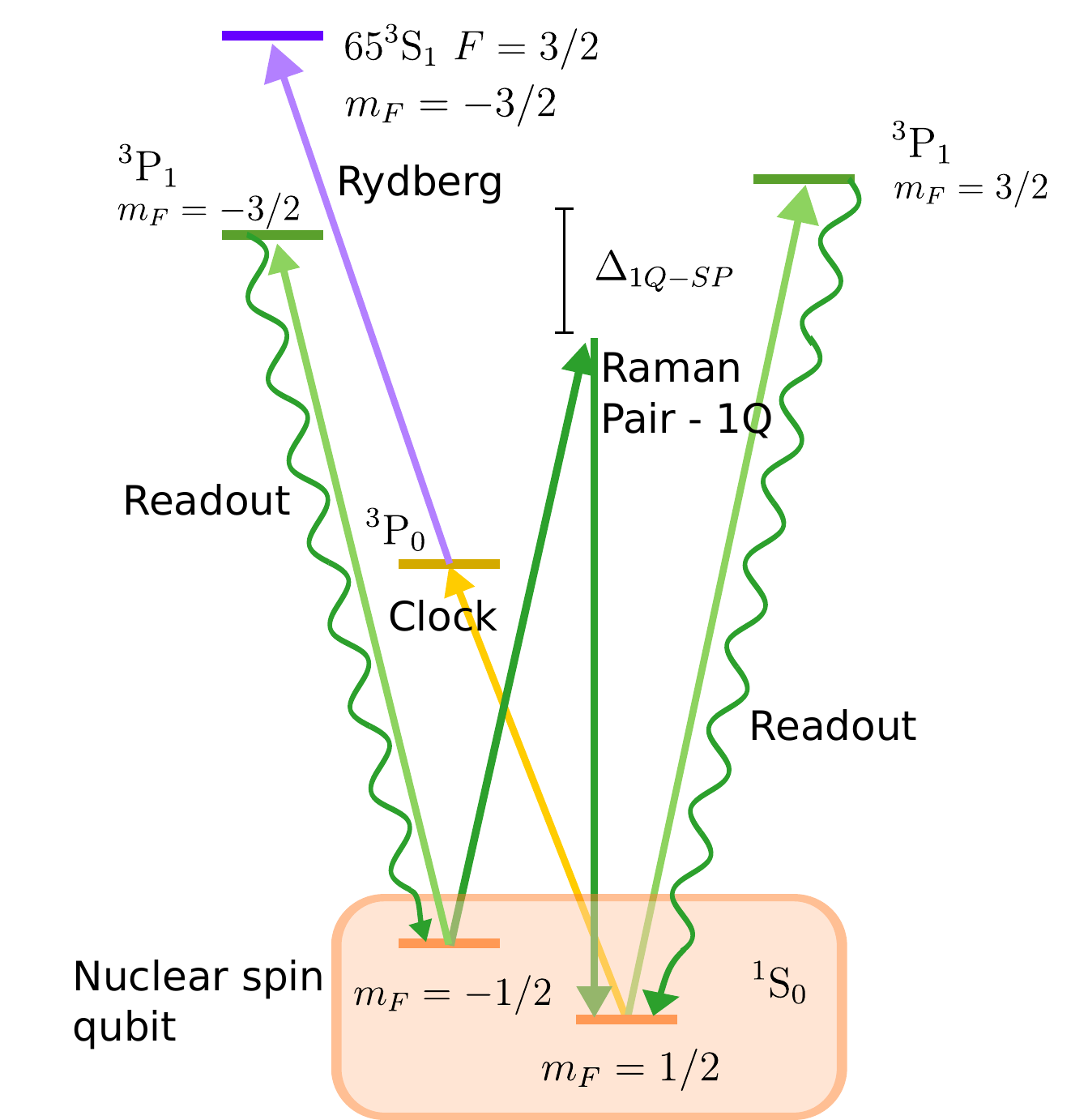}
		\caption{Level diagram indicating the most relevant states for site selective readout, 1Q rotations, clock shelving and Rydberg excitation. Level spacing not to scale}
		\label{fig:levels_detail}
\end{figure}

When atoms are initially loaded in the 483~nm reservoir traps, light assisted collisions and optical pumping addressing the $^1$S$_0\leftrightarrow~^3$P$_1,F=\frac{3}{2}$ transition allows us to prepare single atoms in either of the two ground nuclear spin states. State selective readout is done by addressing the atoms for 5~ms with a single Gaussian beam propagating along the $x$-direction (parallel to the 518~G magnetic field), with waists (w$_y$,w$_z$) = (1.2~mm, 40~$\mu$m) ($1/e^2$ radius), and $\sigma^+$ ($\sigma^-$) polarization to address qubit state $\ket{1}$ ($\ket{0}$), as shown in Figure \ref{fig:fig1}(a) and Figure \ref{fig:levels_detail}. As described in Ref.~\cite{norcia2023midcircuit}, high-fidelity state-selective readout is enabled by a large magnetic field, polarization selectivity, and the use of a magic wavelength. Typical readout has infidelity below 0.2\% and loss below 0.5\%.  

After imaging, atoms are further cooled in the reservoir traps of the interaction zone to facilitate high-fidelity gate operations. In this work we use a variant of gray molasses cooling by addressing the atoms simultaneously with the $\sigma^+$ and $\pi$ polarized 556~nm beams addressing the $^1\textrm{S}_0,m_F=\mp\frac{1}{2}\leftrightarrow ^3\textrm{P}_1, F=\frac{3}{2}, m_F=\frac{1}{2}$ transitions. The $\sigma^+$ beam is generated by the same system as the imaging beam shown in Figure~\ref{fig:fig1}(a), while two $\pi$-polarized beams propagate along the $y$ and $z$ directions \cite{norcia2024iterative,lis2023mid}. These $\pi$-polarized beams are circular Gaussian beams with waists of 150~$\mu$m at the atom plane. The single-photon detuning of these beams is +~2.5~MHz from the excited state, while the frequency difference between the $\sigma^+$ and $\pi$ beams is equal to the qubit frequency of 389~kHz. After a cooling duration of 4~ms, we typically observe $\bar{n}=0.25(10)$, and do not observe significant heating or loss during the transfer of atoms between reservoir and science arrays.

To deterministically load atoms into the IZ, we rearrange atoms with a single 483~nm tweezer created by a pair of crossed acousto-optical modulators (AODs). For this work, we perform CZ gates on either one or two pairs of sites in the IZ. On instances where we load two atomic pairs, these pairs are separated by 12~$\mu$m. After the atoms are loaded, we image, cool and prepare them in state $\ket{0}$ using global addressing. Once the atoms have been initialized, we transfer them to the 460~nm science traps. At this point, we run arbitrary circuits, e.g. GERB, Clifford RB, and RPE calibrations, as described in more detail in the following appendices.

In a typical gate characterization sequence in the IZ we use several readout images with intermediate ground state optical pumping and clock state repumping to determine if each atoms was in state $\ket{0}$, $\ket{1}$, $\ket{c}$, or lost. Data is analyzed in the two-qubit computational basis. We then use some of these images to properly post-select the circuit success based on some conditions, or to identify population recovered by optical pumping and repumping. In this way we can analyze errors arriving from leakage, loss, or anything that affects the qubit subspace. 

\section{Single-Qubit Rotations} \label{1q}

We perform arbitrary local single-qubit rotations using two orthogonally polarized Raman beams whose frequency and position are controlled by two pairs of crossed AODs~\cite{barnes2022assembly}. The Raman beams are combined on a polarizing beam splitter and delivered to the vacuum chamber through a high numerical-aperture objective. At the atom plane, the beams are focused to 1.2~$\mu$m $1/e^2$ radius. One beam is polarized along the magnetic field and the other orthogonal to it. The Raman pair is detuned by $\Delta_{1Q}/(2\pi) = -5$~GHz from the $^3\text{P}_1, F=1/2$ manifold.

Upstream acousto-optical (AOM) and electro-optical (EOM) modulators enable parallel control of single-qubit operations. Calibration of the $X_{\pi/2}$ gate requires calibration of the bare qubit frequency, the differential light-shift on the qubit states due to the addressing lasers, and the pulse area, as well as careful alignment of the Raman beams to the atoms. We calibrate the bare qubit frequency (near 388.9~kHz) through a 100~ms Ramsey spectroscopy experiment in the science traps, where we typically observe changes of $\sim1$~Hz daily. We do not observe significant inhomogeneities in the measured qubit frequency across the IZ sites.

Calibration of the differential light-shifts and pulse area are performed by controlling the relative and common optical powers between the two Raman beams at a single site level. Both calibrations rely on repeated application of pulses, with or without phase changes between them, similar to RPE experiments. Raman beam alignments are realized by mapping an individual beam's AC Stark shift into phase changes in a Ramsey sequence, similar to Ref.~\cite{bluvstein2023logical}. Faster camera-based measurements between 1Q array and the trap arrays are used to correct drifts from its setpoint. Our single-qubit gate performance is not limited by fundamental processes such as intermediate state scattering. Rather, based on RPE and camera-based measurements, we attribute the majority of the 1Q error to quasi-static drifts in the relative position between traps and 1Q beams. Simulations suggest that 100~nm drifts and misalignment can explain typically observed errors.  

GERB sequences for the 2Q gate rely on 1Q operations to randomize the gate of interest over a subset of arbitrary input states. Random Haar distributed rotations are constructed from random angles {$\phi_0,\phi_1,\phi_2$} with distributions weighted such that $R_{\mathrm{rand}} = Z[\phi_0]X_{\pi/2}Z[\phi_1]X_{\pi/2}Z[\phi_2]$ samples each qubit's Hilbert space uniformly. At the end of the GERB sequence, we apply a single pre-computed rotation $R_{\mathrm{f}} = Z[\phi^f_0]X_{\pi/2}Z[\phi^f_1]X_{\pi/2}Z[\phi^f_2]$ to return atoms to the initial qubit state, provided $U$ is perfect. We benchmark our GERB sequence against the 1Q Clifford RB result, by running it with $U = Id$. Each GERB block consist of four $X_{\pi/2}$ gates per atom, so eight per pair. We typically measure 0.32(2)\% infidelity per GERB block (computed on an atomic pair for consistency with 2Q benchmarks), which is consistent with eight times the average gate infidelity measured in the Clifford RB sequence. To remove the 1Q error from other GERB measurements, i.e. when $U$ is not an idle operation, we subtract 0.16(1)\% per $U$ infidelity on each atomic pair.

\section{Clock Operations} \label{clock_app}

The clock beam propagates along the $x$-direction and is elliptically shaped at the atom plane with waists (w$_y$,w$_z$) = (400~$\mu$m, 35~$\mu$m). The large magnetic field and polarization selectivity, see Figure \ref{fig:fig1}(a), suppress other excitation paths, and also suppress differential qubit phase shifts. We operate our clock with Rabi rates $\Omega_{\mathrm{clk}}/(2\pi)$ between 3~kHz and 15~kHz, placing us in the resolved-sideband limit with respect to the radial science trap frequencies of $\omega_{\mathrm{sci}}/(2\pi) \approx 50$~kHz. In this regime, we are sensitive to both the effects of atomic motion and laser frequency noise, which are the main contributors to the shelving error. To mitigate frequency noise seen by the atoms, we stabilize the phase of the delivered clock light via a fiber noise cancellation setup that references the phase of the delivered light to the vacuum chamber~\cite{ma1994fnc}. Optical power delivered to the atoms is also actively stabilized. 

\begin{figure}[htb]
    \includegraphics[width=\columnwidth]{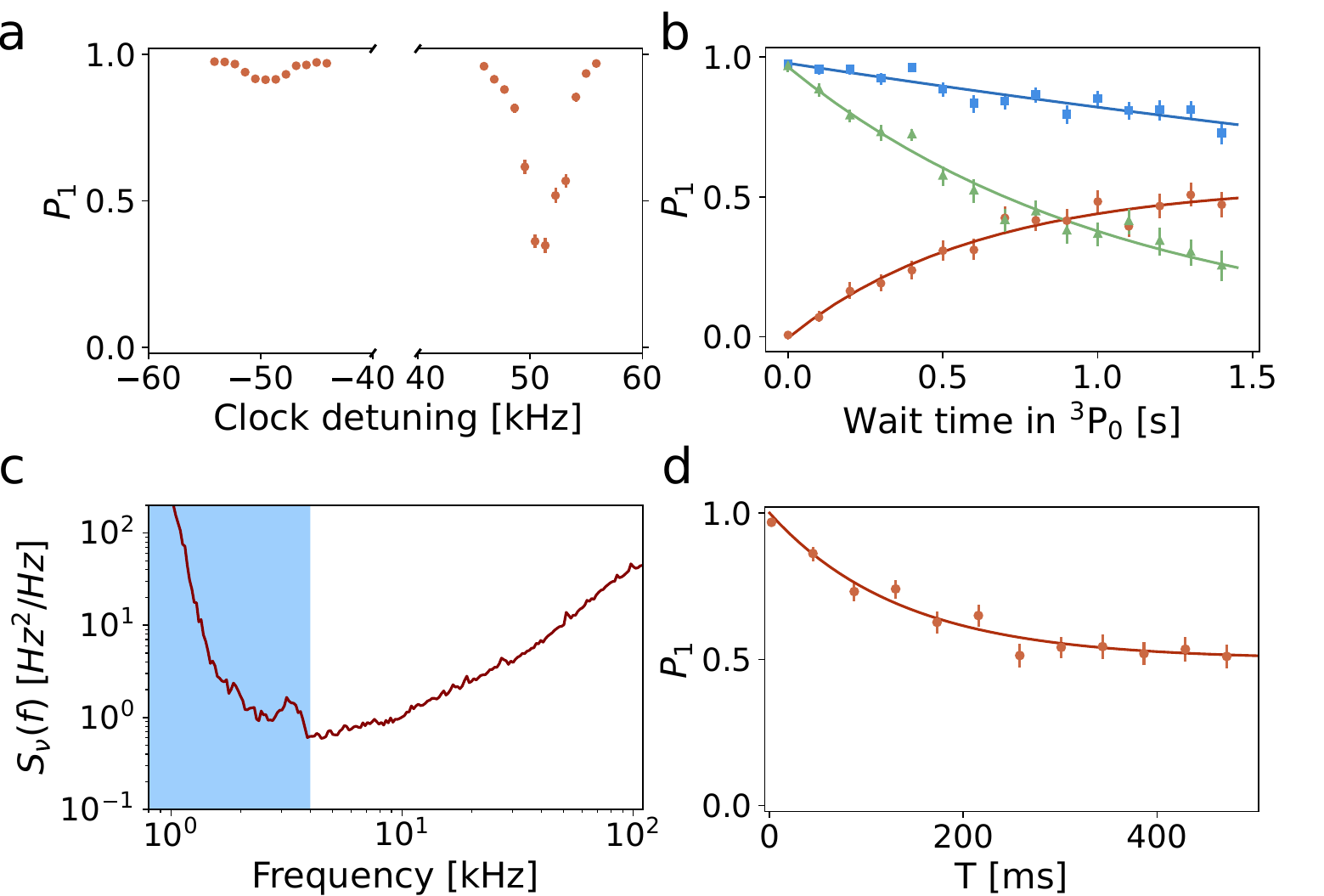}
    \caption{Clock excitation. (a)~Sideband spectroscopy on the clock transition for atoms trapped on the science IZ (average data). (b)~Clock state lifetime. Atoms are held in the science traps (50~kHz trap frequency). We measure population in either ground state (red circles), atom loss (blue squares), and population remaining in the clock state (green triangles), after holding atoms for time $T$. Fitted $T_1$=1.06(5)~s using the $^3\text{P}_0$ survival data (green), while the typical loss via $^3$P$_2$ state and other mechanisms, has a $1/e$ decay time of 5.7(6)~s. (c)~Clock laser frequency noise $S_\nu(f)$ as measured in the 1~km fiber self-heterodyne interferometer at 578~nm. Blue shaded region represents the frequency range where our measurement is overwhelmed by acoustic noise present in the room. (d)~Coherence decay in a spin locking experiment for $\Omega^{\mathrm{Y}}_{\mathrm{SL}}/(2\pi) = 5.5$~kHz. Fit function is $(1+e^{-t(\Gamma_{\mathrm{SL}} + 1/(2T_1))})/2$, where $T_1 = 1.06(5)$~s. Error bars represent $1\sigma$ confidence intervals. }
    \label{fig:clock_sm}
\end{figure}

We characterize the atom temperature using spectroscopy of the motional clock sidebands along $x$, as shown in Figure \ref{fig:clock_sm}(a). At typical temperatures, a single (non-composite) Blackman-shaped pulse with a peak Rabi rate of $2\pi\times7$~kHz shows shelving fidelities slightly above 99\%, measured by repeated application of the clock pulse as in Figure \ref{fig:fig2}(a). We have verified via RPE measurements that residual infidelity is not due to a coherent errors, but rather the Debye-Waller effect from trapped atoms at finite temperature~\cite{wineland_experimental_1998}. Using pulse-area-robust composite pulses improves this shelving fraction, at the price of applying longer pulses with greater sensitivity to both quasi-static and time-varying laser frequency noise.  

The lifetime of ground state atoms in the science or reservoir traps is mostly limited by vacuum and intensity noise on the traps, and is at least 5~s. However, for atoms in the clock state, Raman scattering of the 460~nm light drives population out the $^3$P$_0$ manifold. For science traps with $50$~kHz trap frequency, we have measured $T_1 = 1.06(5)$~s, see Figure~\ref{fig:clock_sm}(b). For the $X^{\mathrm{clk}}_{\pi}$ SCPs pulses used in the main text (260~$\mu$s duration), we estimate that this scattering accounts for an $0.019$\% error per CZ gate averaged over possible pair qubits states (see Appendix~\ref{gate_simulator}). We note that such Raman scattering would act as an idle loss in metastable qubit architectures, posing a challenge to achieve high-fidelity circuit operation at this trap wavelength and depth~\cite{ma_universal_2022,peper_spectroscopy_2024}.

To measure laser frequency noise at 578~nm, we use a scheme similar to the one described in \cite{Kefelian2009FiberInterferometer}. We employ a 1~km SM optical fiber interferometer surrounded by sound-absorbing material. One of the interferometer arms is frequency shifted by a 200~MHz AOM. Laser frequency noise is mapped (via the transfer function of the delay line) into the interferometer beat note at 200~MHz, which is measured with a spectrum analyzer (SignalHound SM200B). A typical measurement is shown in Figure~\ref{fig:clock_sm}(c). At low frequencies, the laser noise is overwhelmed by acoustic noise present in the room (blue shaded region in Figure~\ref{fig:clock_sm}(c)), while we attribute the majority of the noise above 4~kHz to laser frequency noise. 

Atomic spin locking measurements, as in Figure~\ref{fig:fig2}(b), are our best estimation of frequency noise above 1~kHz. A typical spin locking experiments performs a Ramsey sequence that prepares an atom along the Y axis of the optical qubit ($\ket{1} \leftrightarrow \ket{c}$) Bloch sphere, then applies a drive with duration $T$ and Rabi rate $\Omega^{\mathrm{Y}}_{\mathrm{SL}}$ along the Y axis \cite{tsai_benchmarking_2024,finkelstein2024universal}. In the presence of laser frequency noise, the atomic coherence is displaced from the Y axis, such that a final $\pi/2$ pulse along the X axis can not return the population to either of the poles, as shown in Figure \ref{fig:clock_sm}(d). From this data, we determine the $1/e$ decay rate $\Gamma_{\mathrm{SL}}$, taking into account contributions from $T_1$. In the linear response approximation, the decay rate $\Gamma_{\mathrm{SL}}$ is $\Gamma_{\mathrm{SL}} = 2\pi^2S_\nu(\Omega^{\mathrm{Y}}_{\mathrm{SL}}/(2\pi))$, where $S_{\nu}(f)$ is the double-sided laser frequency noise power spectral density \cite{tsai_benchmarking_2024}. We repeat this measurement for different Rabi rates $\Omega^{\mathrm{Y}}_{\mathrm{SL}}$ to reconstruct $S_{\nu}(f)$. 

We find some disagreement between the frequency noise inferred from spin locking measurements and that measured by the fiber interferometer, especially below 10~kHz. This may be due to unaccounted noise sources on the interferometer. For predicting the contribution of laser phase noise, we rely on the frequency PSD $S_\nu(f)$ measured by the spin-locking experiment as an upper bound to the actual laser frequency noise.

To improve data-rate for calibrations involving our clock laser, we perform many experiments in series on the same atoms, which can lead to leakage into the clock state. In order to recycle population from the clock state, we use a repumper on the fast $^3\text{P}_0 \leftrightarrow ^3\text{D}_1, F=\frac{1}{2}$ 1388~nm transition. This beam propagates along the $x$-direction and is linearly polarized. However, about 2\% of the atoms that undergo a repumping cycle decay to the $^3$P$_2$, which is either weakly or not trapped in our tweezers, leading to loss. Therefore, before repumping, we optically pump atoms are in the ground state to $\ket{0}$, apply a high Rabi rate $YXY^{\mathrm{clk}}$ $\pi$-pulse on the $\ket{1}-\ket{c}$ transition, finally apply a resonant 1388~nm pulse to the clock state. This reduces loss significantly, speeding up relevant calibrations. 

\section{Rydberg State} \label{rydberg_state}
\begin{figure}[htb]
    \includegraphics[width=\columnwidth]{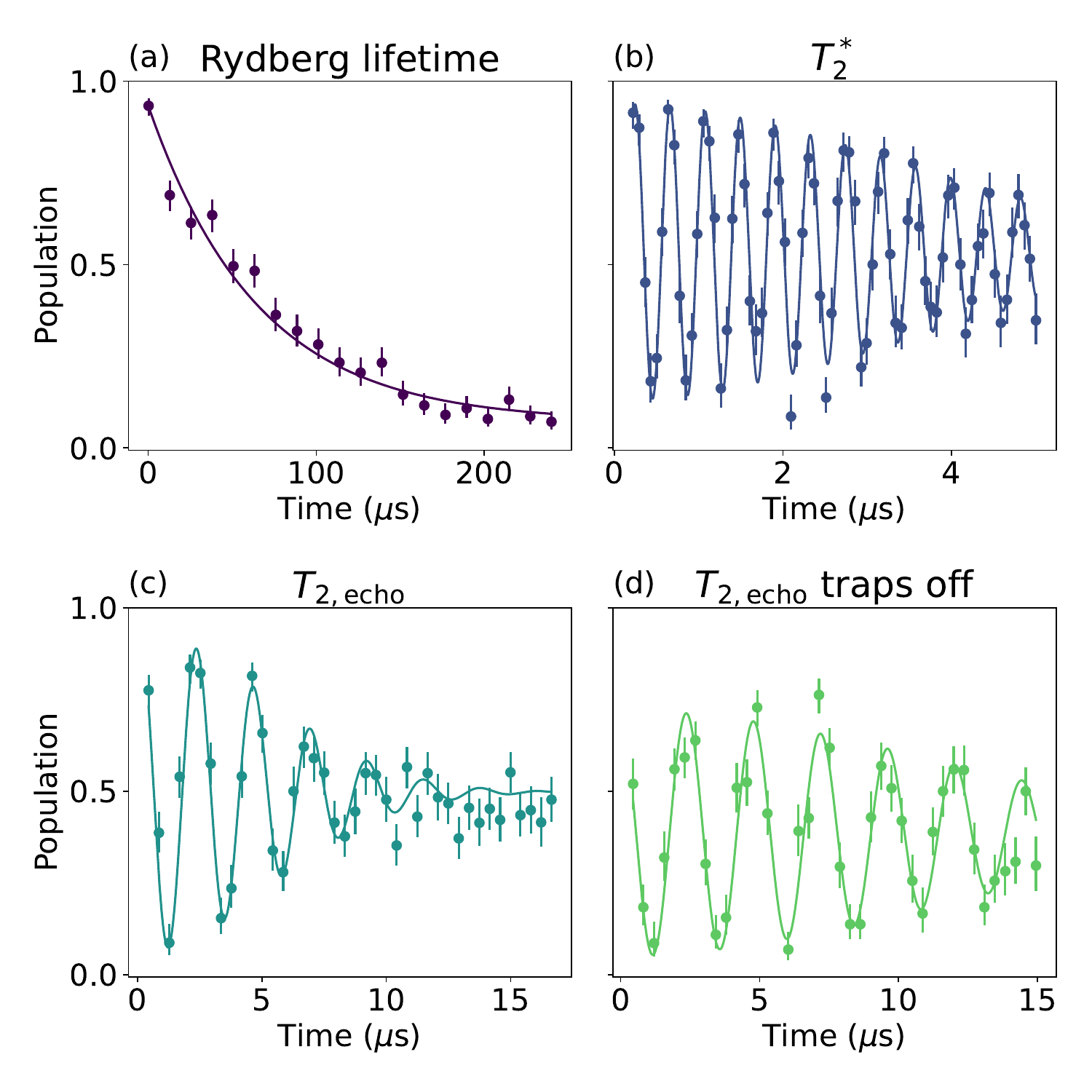}
    \caption{Rydberg state and laser characterization. (a) Rydberg state lifetime measurement, by population return after a $\pi$ pulse into the Rydberg state, variable wait time, and $\pi$ pulse back. Fitted lifetime is 65(3) $\mu$s at a typical trap depth. (b) Clock-Rydberg $T_2^*$ measurement by Ramsey decay, with a Gaussian fit of $T_2^*=3.4(2)~\mu$s. (c) $T_2$ echo decay with traps on, Gaussian fit $T_{2, \text{echo}}=5.1(4)~\mu$s. The $x$-axis represents the total duration of the echo sequence. (d) $T_2$ echo decay with traps dropped, Gaussian fit $T_{2, \text{echo}}=11(2)~\mu$s. Error bars represent $1\sigma$ confidence intervals.}
    \label{fig:ryd_decay}
\end{figure}

Our ultraviolet laser operates at 301.9~nm and is focused down to a approximately circular beam with 18~$\mu$m beam waist, which propagates along the $x$ direction in the interaction zone. We work with the $\ket{r} = \ket{65~^3\text{S}_1,~F=3/2,~m_F=-3/2}$ Rydberg state. Excitation to undesired $m_F$ sublevels is suppressed by both circularly polarized light, and GHz-scale Zeeman splitting from the large magnetic field.

Population and coherence decay times for the  Rydberg state are shown in Figure~\ref{fig:ryd_decay}. As mentioned in the main text, the 460 nm tweezers provide a trapping potential for Rydberg atoms due to the ion core polarizability, and may be kept on or dropped. The lifetime of the Rydberg state is $65(3)~\mu$s with traps on at typical depth, and shows little dependence on the trap depth, so long as it is deep enough to confine the atom effectively. A Gaussian fit to Ramsey contrast decay yields $T_2^*=3.4(2)~\mu$s, which is mostly explained by the UV laser phase noise, dominated by sub-50 kHz frequencies, and also reflects Doppler shifts and other slow detuning errors. An echo sequence gives $T_{2,\text{echo}}=5.1(4)~\mu$s ($11(2)~\mu$s) with traps on (off), as the dephasing due to motion in the trap cannot be echoed effectively at times on the scale of the trap period.

\section{Two-Qubit Gate Calibration} \label{2q_cals}

Entangling gates implemented by transient Rydberg excitation must satisfy several constraints (accrued entangling phase and population left in the Rydberg state for each basis state), and therefore require several degrees of freedom to be tuned. The optimal values may be approximately determined by simulation, but experimental model violation inevitably requires experimental calibration. In special cases the optimal parameters can be inferred one-by-one from a series of experiments, as in the Pichler-Levine gate~\cite{levine_parallel_2019}. Otherwise, the conventional strategy is to parameterize the gate arbitrarily, then optimize the gate performance over 1D scans of each parameter, generally requiring multiple rounds of iteration due to coupling between the parameters.

We address this problem by simulating not just the optimal parameters, but the structure of the optimization landscape. The gate performance is approximated to second order around an optimum as $I\approx I(x_0) + \frac{1}{2}\Delta x_i H_{ij} \Delta x_j$, where $\Delta x_i$ is the difference of the $i$th parameter from its optimal value, and the Hessian matrix can be determined by finite differences of $I$. The Hessian is well-known to be useful in numerical optimization algorithms, but these algorithms are often not robust to experimental noise. We choose instead to diagonalize the Hessian $\boldsymbol{H}=\boldsymbol{Q\Lambda Q^T}$ with $\boldsymbol{\Lambda}$ a diagonal matrix and the columns of $\boldsymbol{Q}$ the (orthogonal) eigenvectors of $\boldsymbol{H}$. Then $I\approx I(x_0) + \frac{1}{2} \sum_i \Lambda_{ii} (\Delta q_i)^2$, where $\Delta q_i = Q_{ji} \Delta x_j$ is the scalar projection of $\boldsymbol{\Delta x}$ along the $i$th eigenvector.

Clearly, the infidelity is now a sum of decoupled functions of the $\Delta q_n$. Thus the $\Delta q_n$ can be optimized independently, scanning over each by adding to $\boldsymbol{x}_\text{initial}$ a multiple of the $n$th eigenvector, $x_j = x_{j,\text{initial}} + \alpha Q_{jn}$, and experimentally optimizing over $\alpha$. The result, as shown in Figure~\ref{fig:eig}, is that the optimum may be reached with a single scan over each eigenvector.

The gate performance function $I$ may be chosen as the average gate infidelity~\cite{jandura2022time} of CZ-X-CZ, appropriate if optimizing in a CZ-GERB sequence which randomizes over input states, or the state infidelity for a single initial state after a sequence with multiple CZ gates, as is done in Figure~\ref{fig:eig}(b). The latter magnifies the sensitivity to entangling phase errors compared to CZ-GERB, but the two yield similar optimal parameters. Either way, we use an echoed sequence to remove the dependence on single qubit phase -- while entangling phase and Rydberg population have nontrivial dependence on the gate parameters, single qubits phases can be easily isolated with an echo, then separately measured and corrected with robust phase estimation and virtual Z gates. 

Experimental imperfections will alter the Hessian, but many (e.g. finite blockade and rise time) can be easily incorporated into the simulation, and in practice the eigenvectors remain reasonably decoupled regardless. With five control parameters such as $A$, $\omega$, $\phi$, $T$, $\delta$, one eigenvector will have a zero eigenvalue, so only 4 tunable parameters are necessary to optimize a Rydberg CPhase gate. We non-dimensionalize all parameters with time units by $\Omega_\mathrm{ryd}$ so that the control eigenvectors can be conveniently scaled to different Rabi rates (however, blockade strength and rise time cannot be easily scaled in experiment, so we pre-compute several Hessians for different ranges of Rabi rates).

\section{Data Analysis} \label{analysis}
\label{analysis}

Our two-qubit gates circuit depth scans repeat a circuit about 20 times and sample 10 different circuits for CZ-GERB and 20 different circuits for 2Q-CRB. The value assigned as the measured probability is the weighted average over the different circuit realizations. Error bars at each circuit depth are assigned as the standard error of the mean among the different realizations.

For circuit depth scans shown in Figures~\ref{fig:fig1}(c), Figure~\ref{fig:fig2}(c) and Figure~\ref{fig:2q_benchmark}, we fit all the decay functions to exponential functions $ap^x+b$, where $x$ represents the depth of the circuit, a and p are fitting parameters, and b is fixed to a predetermined value. Typically $b$ is related to the single atom subspace dimensionality $d$ and the total number of qubits in the measurement basis~\cite{baldwin_subspace_2020}. All infidelities, $1-F$, are calculated from the decay fit as 
\begin{equation}
    1-F = 1 - \frac{(d^2-1)p + 1}{d^2} \approx (1-p)(1-b).
\end{equation}
Confidence intervals on the error rates and fidelities are reported based on the standard deviation errors of the fitting parameters taking into account the actual spread of the experimental data.  

For the 1Q CRB measurement, we consider $b=1/2$, as this is measured in the single atom basis. For the clock-GERB experiments presented in Figure~\ref{fig:fig2}(c) we choose different values of $b$ for the different cases. For the pre-selected fidelity, analyzed in the two-qubit measurement basis, we fix $b=1/9$ because each atom in a pair can be in states $\ket{0}$, $\ket{1}$, or $\ket{c}$ and leakage is larger than loss. For the post-selected fidelity on the qubit-subspace each pair atom can be in states $\ket{0}$ or $\ket{1}$, so we fix $b=1/4$. Finally, the loss measurement represents any other possibility, so we fix $b=0$. 

For the CZ-GERB and 2Q Clifford depth curves in Figure~\ref{fig:2q_benchmark} we make the choice of fixing $b=0$ for the pair survival and pre-selected fidelities, but we fix $b=1/4$ for the qubit-subspace post-selected fidelity as there are four possible basis states for our readout. We set $b=0$ in these cases because atom loss is possible and significant on both curves, and once the circuit fails producing the desired output, the state will not go back to its initial state.

\begin{figure}
    \centering
    \includegraphics[width=\linewidth]{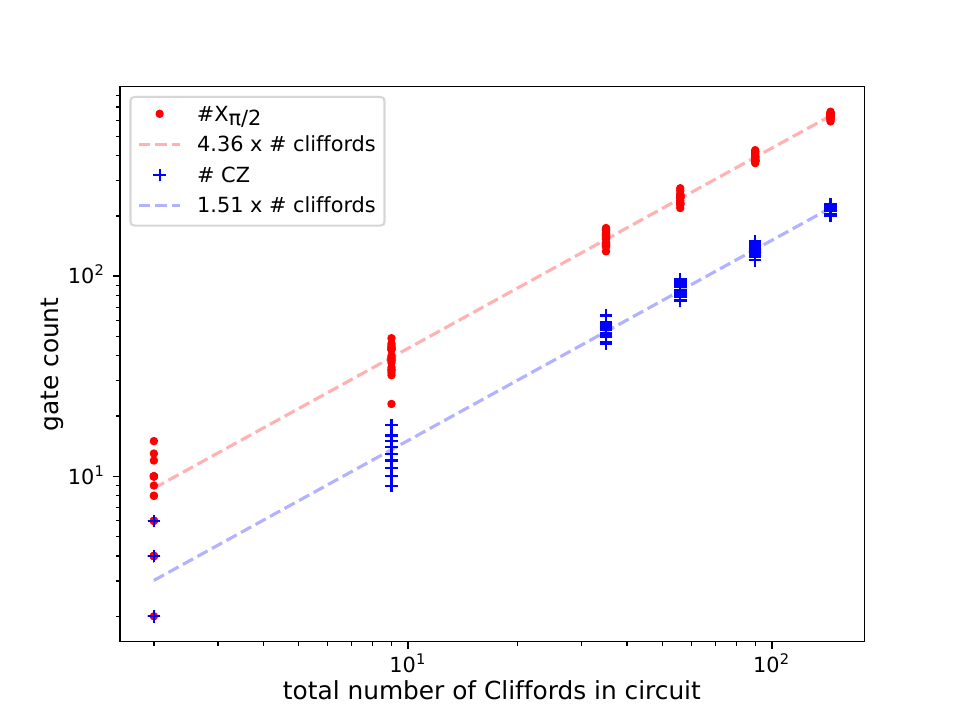}
    \caption{The circuits used for the 2Q Clifford RB experiments, generated as random realizations of circuits with a fixed set of Clifford depths. When these are decomposed into native gates, the number of $X_{\pi/2}$ and CZ gates varies across circuits of the same Clifford depth. For the particular circuits and decomposition into native gates used here, the average number of CZ gates per Clifford gate is 1.51, and the average number of $X_{\pi/2}$ gates is 4.36.}
    \label{fig:gates_per_clifford}
\end{figure}

Isolating the CZ gate error from the 2Q Clifford decay fit requires subtracting out the error incurred by 1Q gates which occur in the sequence. We do this by multiplying the known 1Q CRB error by the average number of 1Q gates in each CZ CRB sequence, and dividing the remaining error by the average number of CZ gates. The average gate numbers are obtained by direct counting of the gates in the circuits used in these experiments, as shown in Figure~\ref{fig:gates_per_clifford}, yielding approximately 4.36 1Q and 1.51 CZ gates per Clifford depth.

\section{Gate Simulator and Error Budget} 
\label{gate_simulator}

Based on atomic experiments, we can estimate the contribution of known error sources. We use an approach where we compute our CZ gate using a super-operator simulator with measured experimental values as inputs. Errors are averaged over time, and represent our understanding of the typical operational state of the machine. This estimation is based on a single gate, rather than a calculation of the fidelity over some specific characterization circuit, i.e. GERB or CRB. 

We separate errors between the clock operation and the UV operation. In the clock, we consider six primary error sources (1) clock frequency drift, (2) fast laser frequency noise, (3) trap scattering out of the clock state to the qubit subspace, (4) finite temperature effects on clock excitation, (5) differential light shift due to the shaped clock pulse between the $\ket{1}$ and $\ket{c}$ state, and (6) loss from the science tweezers. In the UV, we find significant error contribution from the (1) finite Rydberg lifetime, (2) Rydberg decoherence, and (3) pulse repeatability. 

During a calibration cycle, the clock frequency drifts. We estimate a Gaussian distributed static detuning with width $33$~Hz for the clock operation based on experimental measurements of this frequency drift. Since we use a measurement based on detuning several minutes after calibration, this error source is likely an upper bound for the true error due to this effect. This contributes $0.126\%$ infidelity, with $0.007\%$ of this due to leakage.

The clock laser frequency noise is estimated based upon a smoothed fit to experimental data of a spin-lock measurement, and the implied two-sided phase power spectral density from exponential decays in the spin locking experiment shown in Figure \ref{fig:fig2}(b). This fit goes down to $10$~Hz, and thus is distinct from the slow time scale drift. For a given realization of the phase noise power spectral density $S_\phi$, we sample the laser phase as in~\cite{Jiang2023}.
\begin{align}
    \phi(t) = 2 \sum_k \sqrt{S_\phi(f_k) \Delta f_k} \cos \left( 2 \pi f_k t + \phi_k \right),
\end{align}
with $\phi_k$ chosen randomly and $\Delta f$ some small step in frequency space. This phase adds simply to the desired phase of the pulses. The clock phase noise contributes $0.116 \%$ to the infidelity and $0.088 \%$ to the leakage rate.

In order to model the effects of atom temperature on the clock shelving fraction, we consider only the modification to the resonant carrier Rabi rate due to the Debye-Waller effect. Here, the Rabi rate for the $n^{th}$ motional eigenstate along the direction of the driving laser is reduced by $e^{-\eta^2 / 2} L_n(\eta^2)$, with $\eta \approx 0.26$ the Lamb-Dicke parameter and $L_n$ the $n^{th}$ Laguerre polynomial. In order to approximate the pulse area calibration procedure which takes place at this same temperature, we set the thermally average Rabi rate to give a perfect pulse area for the given pulse duration. Thus, colder than average atoms will be over-rotated while warmer than average atoms will be under-rotated. We choose $\bar n = 0.25$, and sample the motional eigenstate independently for both atoms. Temperature effects in the clock cause $0.013  \%$ infidelity entirely in the form of leakage. 

Raman scattering of off resonant trap light from the clock state is a known error source. In our nominal traps, the clock state has a lifetime of $1.06$~s. In order to simplify the analysis here, we model the leakage scattering as having a branching ratio of $50 \% $ to each atomic ground state, since the lifetime of the main intermediate state $^3P_1$ in the decay pathway is short compared to the clock Rabi rate. This contributes $0.019 \%$ to infidelity, with $0.008\%$ leaking back to the clock state. The observed $5$~s $1/e$ atom loss contributes to $0.005\%$ leakage.  

Additionally, due to the AC Stark shifts from far off-resonant states, the clock operation shifts the differential frequency between $\ket{1}$ and $\ket{c}$ by an average of $125$~Hz over the course of the clock pulse. We calibrate the detuning so that the time-averaged detuning is $0$, but the light shift varies over the course of the pulse, and the shaped clock pulse will only be resonant at its average Rabi rate. This contributes $0.009\%$ error, entirely as leakage since the average detuning is $0$.

Turning to UV errors, we model the Rydberg lifetime as $65~\mu s$, where an atom lost from the Rydberg state is assumed to be permanently lost. This leads to a loss probability of $0.075 \%$ per pair during the CZ gate. Assuming the measured $T_2^*$ results from quasi-static detuning errors, the root mean square (rms) detuning can be calculated as $2\pi\times\Delta_\text{rms}=\frac{1}{T_2^*}$. Simulation of the time-optimal gate with a static detuning error yields an infidelity $2.9 (2\pi\Delta/\Omega)^2$, for a total infidelity $2.9 / (T_2^* \Omega)^2 = 0.007\%$. Similarly,we expect a small coherent population left in the Rydberg state, that we treat as loss, which contributes $0.001\%$ to such error. We measure atom loss of 0.11(3)\% and 0.09(1)\% for GERB and CRB after accounting for clock loss (see Table I). The small discrepancy between the predictions of our model and RB measurements can potentially be explained by gate parameter drift or excess decay of the entangled Rydberg state.

We additionally model a $0.4\%$ variation in the time optimal pulse area, constant over the duration of the UV gate as measured in the experiment. This contributes error at $0.007\%$, mostly due to the $0.005\%$ loss. 

In general, when calculating the fidelity for an arbitrary CZ gate, we are free to choose a single-qubit phase as we see fit. We choose the single qubit phase which minimizes each error independently, in order to avoid over estimating coherent errors. We finally recalibrate the parameters of the time-optimal gate for our finite Rydberg blockade of $160$~MHz.

\begin{table}[htb]
    \centering
    \begin{tabular}{|p{3.5cm}|p{2cm}|p{2cm}|}
    \hline 
    Error source & Infidelity & Leakage/Loss \\
    \hline
    Clock detuning & 0.126\% & 0.007\% \\ 
    \hline 
    Clock frequency & 0.116\% & 0.088\% \\
    \hline
    Clock temperature & 0.013\% & 0.013\% \\
    \hline
    Clock scattering & 0.019\% & 0.008\% \\
    \hline
    Clock loss & 0.005\% & 0.005\% \\
    \hline
    Clock light shift & 0.009\% & 0.009\% \\
    \hline
    UV $T_1$ & 0.075\% & 0.075\% \\
    \hline
    UV $T_2^*$ & 0.007\% & 0.001\% \\
    \hline
    UV repeatability & 0.007\% & 0.005\% \\
    \hline
    Totals & 0.375\% & 0.211\% \\
    \hline

    \end{tabular}
    \label{tab:sim_infidelities}
    \caption{Summary of simulated error sources in our CZ gate, as compared to fig.~\ref{fig:error_budget}.}
    \label{tab:sim_infidelities}
\end{table}

\begin{figure}[h]
    \centering
    \includegraphics[width=\linewidth]{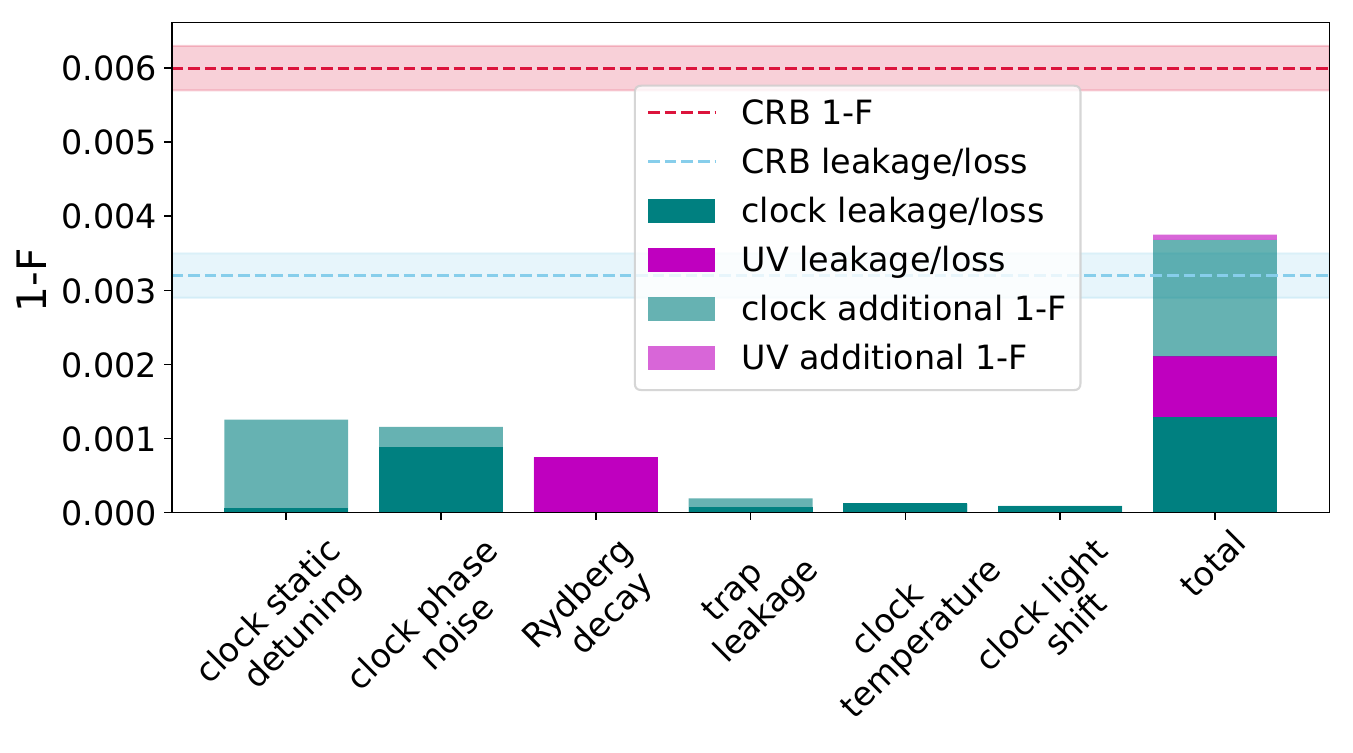}
    \caption{Error budget based on simulation and analytic calculations. The darker shaded region in each bar represents the portion of infidelity due to leakage and loss. We include our best understanding of all input error sources from \emph{ab initio} experimental measurements. }
    \label{fig:error_budget}
\end{figure}

We summarize the contributions of each error source with $1-F > 10^{-4}$ in Figure ~\ref{fig:error_budget} and also in Table~\ref{tab:sim_infidelities}. In the figure, the height of each bar represents the decrease in average gate fidelity due to this error. The darker portion of the bar gives the probability for either atom to be outside of the qubit subspace following the CZ gate, i.e. at least one atom is either leaked or lost. The lighter portion alone is the portion of the error not due to leakage or loss. The bars are colored according to whether they occur on the clock operation (green) or the UV operation (magenta). Along with these error sources, we show the experimentally measured CZ CRB and GERB infidelities, as well as the numerically modeled totals for both leakage and loss, and the decrease in fidelity $\Delta F$. This error budget predicts an infidelity of $0.375\%$ with at least one atom leaked or lost $0.211\%$ of the time. 

When realizing a CZ-GERB experiment, effects such as single-qubit phase errors are not relevant due to the global echo present in each GERB block. Therefore, we expect most of the CZ-GERB error to be caused by leakage, loss and any gate parameter calibration errors. For other metrics such as the Clifford RB ones, additional errors that are not echoed, for example that affect the single-qubit phase, contribute to the measurement.

Regarding the disagreement between the simulated error budget and experimentally measured errors, we have identified some potential error sources that will be subjected to further investigation. We decline to include errors in the error budget which are not experimentally well-characterized. For example, we observed a $\sim 30$\% larger clock leakage and loss than what we expected due to clock related errors, which are transferred into the CZ gate. We have also observed larger loss than the one predicted by the single atom Rydberg $T_1$ measurements with the 460~nm traps on during the measurement, but the Rydberg $T_2$ and $T^*_2$ measurements are fully explained by the UV laser frequency noise. Although experimentally the time between consecutive CZ gates is always much larger than the Rydberg $T_1$, the dynamics of these populations over long circuits are also not considered in this gate simulator, which could impact deeper circuits.

We know of other potential effects that can contribute to this error budget, some are relatively small, i.e. clock laser intensity noise, Doppler effects, finite Rydberg blockade. However, other effects are less well-known, for example the effect of atom motion in an imperfect trapping potential when probed by a Doppler-sensitive operation such as our clock shelving, and a complete understanding of how Rydberg pair states affects the CZ gate under practical operational conditions that involves multiple lasers and background fields. Future work will investigate the impact of these additional effects on our CZ gates.

\bibliography{bib}
\end{document}